\documentclass[a4paper,10pt,twocolumn]{article}
\usepackage{graphicx}
\usepackage{bm}
\usepackage{cite}
\usepackage{authblk}
\usepackage{hyperref}

\usepackage[top=1.5cm,bottom=1.5cm,left=1.5cm,right=1.5cm]{geometry}
\usepackage[mathlines]{lineno}

\title{Experimental characterization of the transition to coherence collapse in a semiconductor laser with optical feedback}
\author[1,2]{M.~Panozzo}
\author[1]{C.~Quintero-Quiroz}
\author[1]{J.~Tiana-Alsina}
\author[1]{M.C.~Torrent}
\author[1]{C.~Masoller.}
\affil[1]{Universitat~Polit\`ecnica~de~Catalunya, Departament~de~F\'isica, Colom~11, 08222~Terrassa, Barcelona, Spain.}
\affil[2]{Universit\`a~degli~Studi~di~Padova, 
Via~VIII~Febbraio~2, 35122~Padova , Italy.}
\date{}
\providecommand{\keywords}[1]{\textbf{\textit{Keywords---}} #1}
\begin{document}

\twocolumn[
\maketitle
\begin{abstract}
Semiconductor lasers with time-delayed optical feedback display a wide range of dynamical regimes, which have found various practical applications. They also provide excellent testbeds for data analysis tools for characterizing complex signals. Recently, several of us have analyzed experimental intensity time-traces and quantitatively identified the onset of different dynamical regimes, as the laser current increases. Specifically, we identified the onset of low-frequency fluctuations (LFFs), where the laser intensity displays abrupt dropouts, and the onset of coherence collapse (CC), where the intensity fluctuations are highly irregular. Here we map these regimes when both, the laser current and the feedback strength vary. We show that the shape of the distribution of intensity fluctuations (characterized by the standard deviation, the skewness and the kurtosis) allows to distinguish among noise, LFFs and CC, and to quantitatively determine (in spite of the gradual nature of the transitions) the boundaries of the three regimes. Ordinal analysis of the inter-dropout time intervals consistently identifies the three regimes occurring in the same parameter regions as the analysis of the intensity distribution. Simulations of the well-known time-delayed Lang-Kobayashi model are in good qualitative agreement with the observations.
\end{abstract}

]
\begin{quotation}
Complex dynamical systems often show sudden or gradual transitions between distinct dynamical regimes. Examples include desertification transitions as precipitation decreases, population extinctions when food becomes scarce, bistable visual perceptions, sleep--wake transitions and epileptic seizures. The last years have seen a rapid increase of data--driven approaches aimed at gaining deeper insights into the detection and prediction of regime transitions. Here we study an experimental optical system (a diode laser with time-delayed optical feedback) that displays a gradual transition from noise to chaos, as the control parameters (the laser current and the feedback strength) are varied. We demonstrate that statistical and symbolic data analysis tools clearly identify the onset of different regimes, despite the gradual nature of the transitions. These tools can be valuable for analyzing regime transitions in many complex systems.
\end{quotation}
\section{\label{sec:intro}Introduction}

Semiconductor lasers with optical feedback are important practical devices and their chaotic output has found many applications \cite{uchida_nat_phot_2008,donati_2012,uchida_chaos_2012,ingo_nat_comm_2013,marc_nat_phot_2015,rontani_sci_rep_2016}. In addition, the dynamics generated by optical feedback is relevant from the complex systems' perspective. The laser dynamics is nonlinear (due to light-matter interaction in the laser cavity), stochastic (due to spontaneous emission noise, thermal and electrical noise) and high-dimensional (due to the feedback delay time) \cite{masoller_chaos_1997,parlitz_pre_1998,vicente_jqe_2005,yanchuk_prl_2014}. Thus, the output intensity emitted by a semiconductor laser with optical feedback is very attractive for testing a variety of data analysis tools \cite{tiana_pra_2010,soriano_jqe_2011,zunino_jqe_2011,nico_pre_2011,aragoneses_2014,deb_oe_2014,porte_autocorrel_pre_2014,porte_similarity_pra_2014,masoller_njp_2015,masoller_plos_2016}. 

Several of us have recently investigated experimentally the transition, as the pump current is increased, from optical noise to chaos \cite{quintero_sci_rep_2016}. The onset of the chaotic regime is accompanied by a drastic enhancement of the laser line-width and relative intensity noise \cite{lenstra_1985,feedback_regimes_1986,shore_ol_1994,ingo_coex_1998}; however, in the time domain the transition is gradual: starting at the lasing threshold (optical noise), as the laser pump current is increased there is a continuous transition during which the intensity starts displaying dropouts, which become increasingly regular (a regime known as low-frequency fluctuations --LFFs), and then gradually become faster and irregular \cite{video}. When the dropouts are too fast to be distinguished as individual events, the dynamics is referred to as fully developed coherence collapse (CC). In \cite{quintero_sci_rep_2016} we have shown that three diagnostic tools (the standard deviation of the intensity time series, the number of threshold-crossing events, and symbolic ordinal analysis\cite{bandt2002permutation}) provide complementary information about the properties of these regimes and allow to quantitatively distinguish between optical noise, LFFs and CC. Remarkably, the three diagnostic tools consistently identified the noise, LFFs and CC regimes occurring in the same intervals of the experimental control parameter, the laser pump current. 

An open question is how the feedback strength affects the properties of the noise--LFF and LFF--CC transitions, and in particular, how the transition points depend on the feedback strength. A recent study has mapped these regimes when the feedback strength and the laser current are varied~\cite{uchida_2017}. The study was performed in the so-called short cavity regime, when the external cavity is short enough to yield a feedback delay time, $\tau$, that is shorter than the natural laser relaxation oscillation period, $T_{ro}$~\cite{ingo_2001,marc_2004,pepe_2013}. Here, as in~\cite{quintero_sci_rep_2016}, we consider the long cavity regime in which $\tau>>T_{ro}$. We experimentally record the intensity dynamics with different feedback levels and show that both, the shape of the intensity distribution (characterized by the standard deviation, the skewness and the kurtosis) and the temporal correlations among consecutive inter-dropout intervals (characterized by using symbolic ordinal analysis) clearly distinguish, in the parameter space (pump current, feedback strength), the regions of noise, LFFs and CC. Simulations of the time-delayed Lang-Kobayashi model \cite{lk_model} are in good qualitative agreement with the experimental observations.  

This paper is organized as follows: Sec.~\ref{sec:exp} presents the experimental setup; Sec.~\ref{sec:methods} presents the diagnostic tools and Sec.~\ref{sec:exp_results} presents the experimental results. Sec.~\ref{sec:numerical} presents the Lang-Kobayashi model and the numerical results. Finally, Sec.~\ref{sec:con} summarizes the conclusions.

\section{\label{sec:exp}Experimental setup}

The experimental setup is as in Ref.~\cite{quintero_sci_rep_2016}. A 685 nm semiconductor laser (AlGaInP multi-quantum well HL6750MG), with solitary threshold current $\mathrm{I_{th,sol}}=26.74 $~mA has part of its output intensity fed back to the laser cavity by a mirror. The length of the external cavity is 70~cm which gives a feedback delay time of about 5~ns. The laser temperature and current were stabilized using a combi-controller (Thorlabs ITC501) with an accuracy of 0.01 C and 0.01 mA, respectively. During the experiments the temperature was set to T = 18 C. 

A 90/10 beam-splitter in the external cavity sends light to a photo-detector (DET10A/M Silica based photodetector) that is connected to an amplifier (FEMTO HSA-Y-2-40) and a 1 GHz digital storage oscilloscope (Agilent Technologies Infiniium DSO9104A). 

A set of 9 neutral density filters (NDFs with optical densities OD=0, 0.2, 0.4, 0.6, 0.8, 1, 1.2, 1.5, 2 [dB]) in the external cavity allowed to vary the strength of the feedback from 18.15\% to 2.6\% threshold reduction. A LabVIEW program was used to control the experiment. The pump current was varied from 20.50 mA to 34 mA in 92 steps. For each set of (pump current, NDF	OD), $N=10^7$ data points were recorded with sampling rate of 2 GS/s (temporal resolution of 0.5 ns), allowing an observation time of 5 ms. It it important to note that the intensity dropouts are the envelope of fast, pico-second pulses, which can be detected by using a much faster detection system~\cite{ingo_prl_1996}. 

\section{\label{sec:methods}Diagnostic tools}

The first diagnostic tool is based in quantifying changes in the shape of the intensity probability distribution function (pdf) as the laser current or the feedback strength are varied. We use the standard deviation, $\sigma$, the skewness, $S$, and the kurtosis, $K$, to distinguish among noise, LFFs and CC. While $\sigma$ quantifies the width of the intensity pdf, $S$ and $K$ measure the degree of asymmetry and the degree of flatness, respectively.

We remark that the amplifier used in the experimental detection system removes the mean value of the intensity time series.  Thus, for each intensity time series, $\{I_1,\dots, I_i, \dots\, I_N\}$ with $\left<I_i\right>=0$, $\sigma$, $S$, and $K$ are calculated as
\begin{eqnarray}
\sigma^2 &=&\frac{\sum_i I_i^2}{N-1}\\
S&=&\frac{\sum_i I_i^3}{(N-1)\sigma^3}\\
K&=&\frac{\sum_i I_i^4}{(N-1)\sigma^4}.
\label{definitions}
\end{eqnarray}
    
The second diagnostic tool is based in the symbolic analysis of the time intervals, $\{\Delta T_1, \dots, \Delta T_{i} = t_i-t_{i-1}, \dots\}$, between consecutive intensity dropouts (inter-dropout intervals, IDIs). To define the timing of the dropouts two thresholds are used~\cite{quintero_sci_rep_2016}: $t_i$ is defined as the time when the intensity falls below a predefined detection threshold; then, the intensity has to grow above a second threshold, $\left<I_i\right>=0$, before the next dropout can be detected. This second threshold is needed in order to avoid detecting as dropouts the fluctuations that occur during the recovery process. 

Because the amplitude of the intensity fluctuations (and in particular, the depth of the intensity dropouts) varies with the control parameters (pump current, feedback strength), in order to use a uniform criterion for defining ``dropouts'', the detection threshold is chosen to be proportional to $\sigma$. To select an appropriate detection threshold we need to take into account that too low thresholds have the drawback of missing many dropouts (the intensity falls, but not deep enough to cross the threshold), while too high thresholds detect not only dropouts, but also, many noisy fluctuations. A detailed analysis of the influence of the threshold was presented in~\cite{quintero_sci_rep_2016}. Here we use a threshold $-1.5\sigma$ because it provides a good compromise between filtering noise while detecting a sufficiently large number of dropouts: with this threshold the intensity time series contains more than $10^4$ dropouts, depending on the laser current and feedback strength [see Fig. 1(b)]. 

The resulting sequence of inter-dropout intervals is analyzed by using the ordinal methodology \cite{bandt2002permutation}, which consists of transforming the sequence of time intervals into a sequence of ordinal patterns (OPs), defined by considering the relative length of $D$ consecutive intervals. For $D=3$ there are six OPs: 

$\Delta T_{i} < \Delta T_{i+1} <  \Delta T_{i+2}$ gives $012$, 

$\Delta T_{i} < \Delta T_{i+2} <  \Delta T_{i+1}$ gives $021$, 

$\Delta T_{i+1} < \Delta T_{i} <  \Delta T_{i+2}$ gives $102$,

...

$\Delta T_{i+2} < \Delta T_{i+1} < \Delta T_{i}$ gives $210$. 

In this work we only consider $D=3$ OPs because it is sufficient to detect a certain degree of determinism in the IDI sequence. As $D$ increases the number of OPs increases as $D!$. Because the IDI sequence is noisy, very long sequences of dropouts are needed in order to detect patterns whose probability significantly deviates from $1/D!$~\cite{gray_region}, i.e., patterns that appear in the sequence more often (or less often) than expected in the null hypotesis that all patterns are equally probable.

This ordinal approach has the drawback that two very different types of intensity dynamics can give a similar set of ordinal probabilities. First, when the intensity dropouts are fully irregular, the IDI time series is also fully irregular; in this case the ordinal sequence will be random and the six probabilities will be within the interval of values which are consistent with 1/6. On the other hand, when the intensity dropouts are almost periodic, the IDIs will fluctuate around a well-defined mean value (the period of the dropouts), and if the fluctuations are uncorrelated, the six patterns will also be expressed in the ordinal sequence with equal probability. 

\begin{figure}
\centering
\includegraphics[width=\columnwidth]{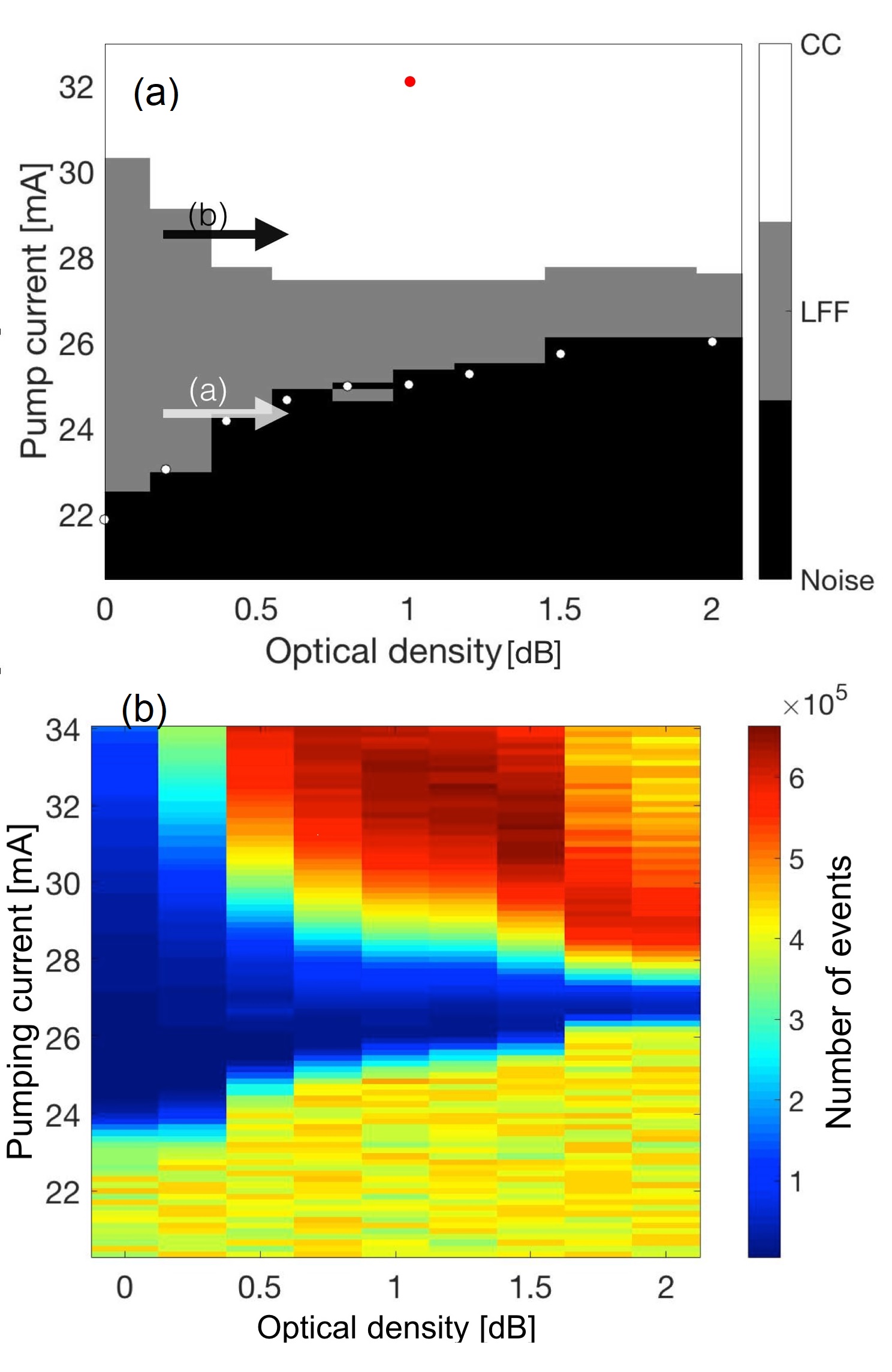}
\caption{ (a) Experimental map representing the occurrence of different dynamical regimes (noisy intensity fluctuations, low frequency fluctuations --LFFs, and coherence collapse --CC) in the parameter space (laser current in mA, feedback strength in units of the optical density of the neutral density filter). The arrows and the red dot indicate the parameters used in Fig.~\ref{fig:2} and the white dots indicate the threshold of the laser with feedback (measured from the LI curve). (b) Number of dropouts in each time series: number of times the intensity falls below -1.5$\sigma$ (with $\sigma$ being the standard deviation of the intensity pdf).}
\label{fig:1}
\end{figure}

\begin{figure*}
  \centering
  \includegraphics[width=.75\textwidth]{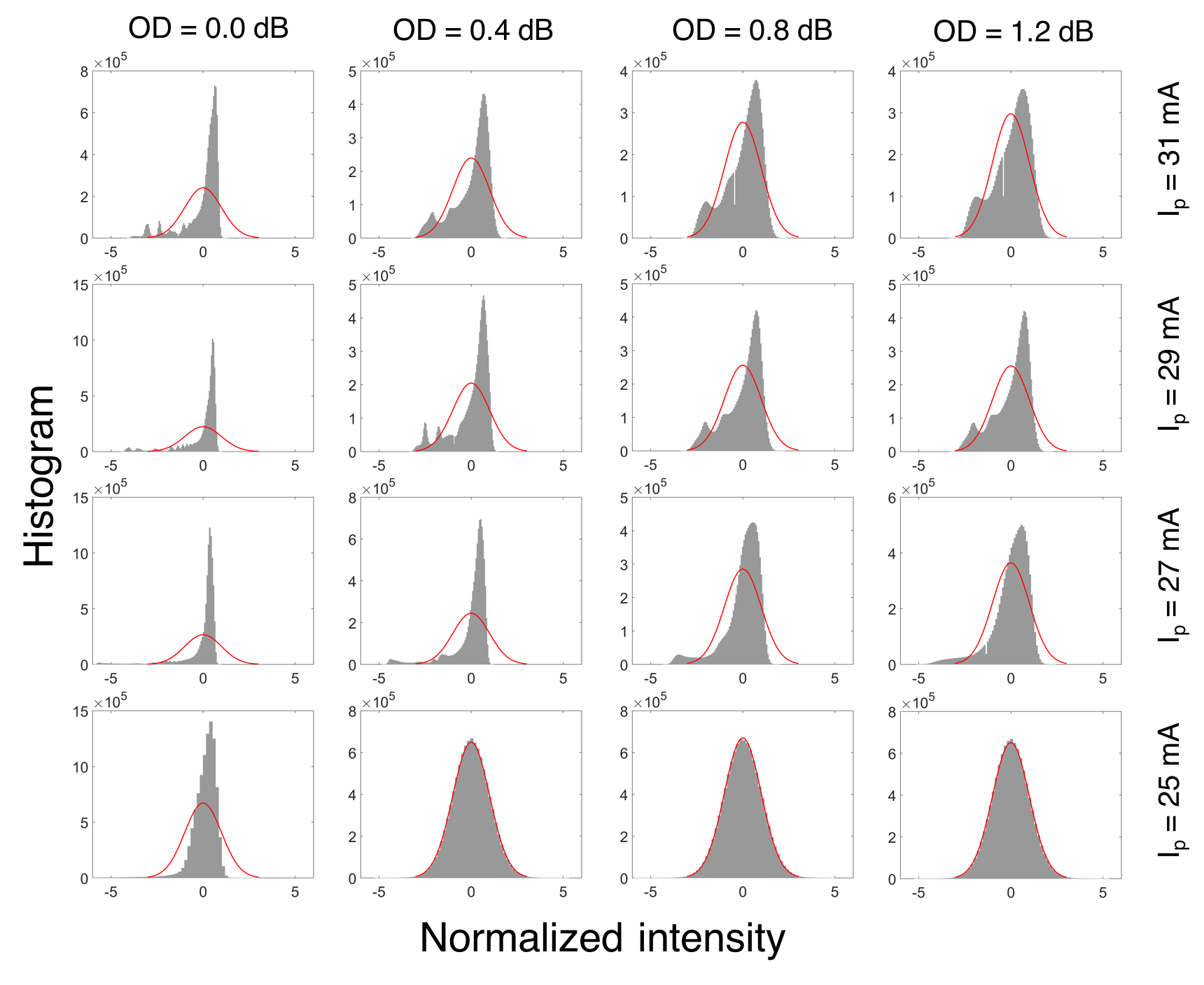}
  \caption{Experimental intensity pdfs as a function of the laser current and feedback strength. The solid lines indicate the fit to a Gaussian distribution.}
  \label{fig:histograms}
\end{figure*}

\begin{figure}
\label{fig:2}
\centering
\includegraphics[width=.5\textwidth]{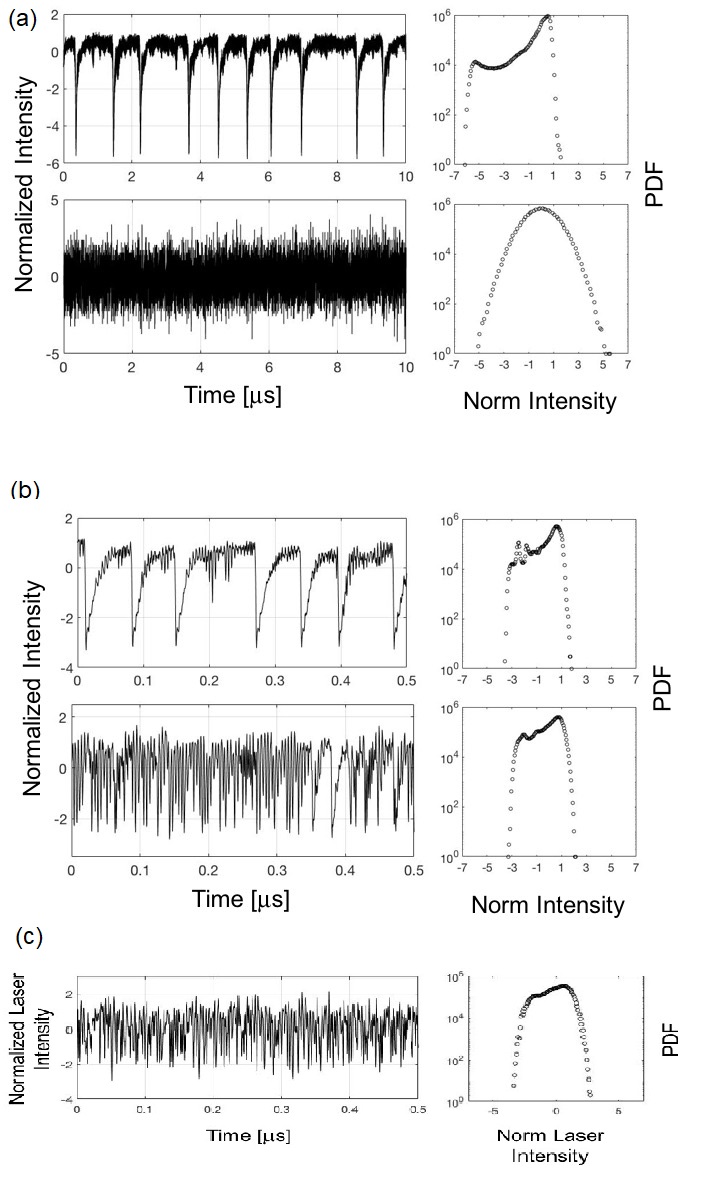}
\caption{ Experimental intensity time series (normalized to $\left<I_i\right>=0$ and $\sigma=1$) and histograms for two pump currents and two feedback strengths. (a) noisy and LFF regimes (pump current 24.56 mA and NDF OD=0.2 and 0.6 [dB], respectively); (b) LFF and CC regimes (pump current 29.51 mA and NDF OD=0.2 and 0.6 [dB] respectively). These parameters are indicated with arrows in Fig.~\ref{fig:1}(a). In (c) fully developed coherence collapse is observed for pump current 32.02 mA and NDF OD=1 [dB]. This set of parameters is indicated with a dot in Fig.~\ref{fig:1}(a)}.
\end{figure}

\section{\label{sec:exp_results} Experimental results}

Figure~\ref{fig:1}(a) displays the parameter regions where the noise, LFF and CC regimes are identified in the plane (pump current, feedback strength), and well defined boundaries are observed. To obtain this map, the following criteria were used to classify each time series in one of the three regimes:

1) We identify noisy emission by comparing the intensity distribution with a Gaussian distribution: when the skewness of the intensity pdf is below a threshold (chosen equal to the maximum skewness observed for the intensity pdf of the solitary laser) and the kurtosis of the intensity pdf is within the interval [3--3.3] (i.e., it differs up to 10\% from the kurtosis of a Gaussian), then the intensity pdf is considered consistent with a Gaussian pdf and the time series is classified as noise. 

2) If the first criterion is not fulfilled, then, as in \cite{quintero_sci_rep_2016}, the variation of $\sigma$ with the laser current (keeping the feedback strength constant) is used to distinguish between LFF and CC:  if $d\sigma/d\mathrm{I}>0$, then the time series is classified as LFFs, else, as CC. 

The resulting map is robust with respect to small variations of the skewness or kurtosis thresholds, and is in good agreement with previous observations~\cite{ingo_coex_1998,deb_oe_2014}. As shown in Fig.~\ref{fig:histograms}, in the region identified as noisy emission the intensity distribution is well fitted by a Gaussian, while in the regions identified as LFFs or CC, it significantly deviates from a Gaussian.

Figure~\ref{fig:2} displays examples of the intensity dynamics in the different regions, and the corresponding distributions in log scale [the parameters are indicated with arrows and a dot in Fig.~\ref{fig:1}(a)]. Clear differences are observed both, in the dynamics and in the shape of the distribution, which are captured by the classification criteria described above. 

\begin{figure}
\centering
\includegraphics[width=\columnwidth]{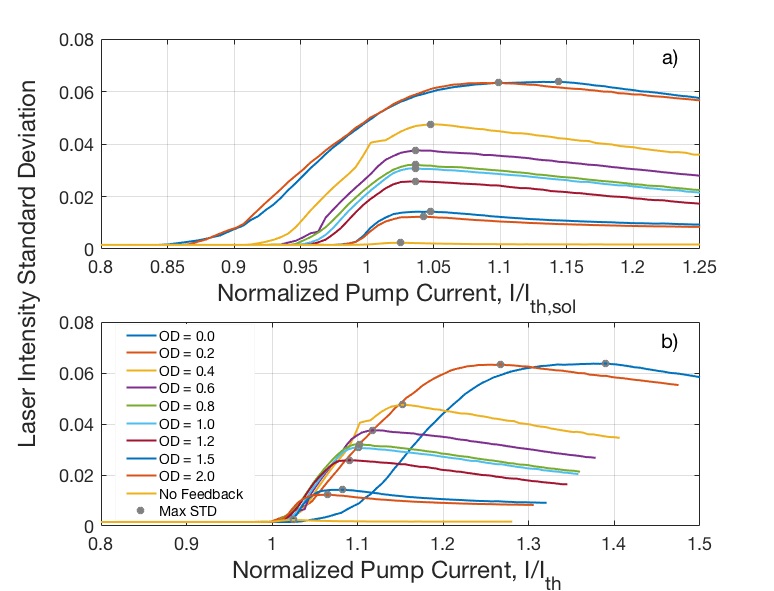}\\
\label{fig:3}
\caption{(a) Standard deviation of the experimental intensity time series vs. the laser current, normalized to the solitary threshold, $\mathrm{I_{th,sol}}$, for various feedback strengths. The dots indicate where $\sigma$ is maximum, which identifies the onset of the LFF--CC transition. We note that at fixed current, $\sigma$ grows with the feedback strength (except for the strongest feedback, OD=0). In (b) the laser current is normalized to the threshold of the laser with feedback, $\mathrm{I_{th}}$. Here we note that, except for the strongest feedback, the shape of the curve $\sigma$ vs. $\mathrm{I}/\mathrm{I_{th}}$ is similar for all feedback strengths.} 
\end{figure}

In Fig.~\ref{fig:3} we analyze the variation of $\sigma$ with the laser current, for various feedback strengths. In~\cite{quintero_sci_rep_2016} the current value for which $\sigma$ is maximum (indicated with a dot) was identified as the onset of the LFF--CC transition. We note that, for strong feedback, the LFF-CC transition moves to higher pump currents as the feedback increases; in contrast, for weak feedback (OD between 0.6 and 1.2) the pump current at which the LFF-CC transition occurs remains nearly constant as the feedback increases.

In Fig.~\ref{fig:3}(a) we also note that, at fixed current, $\sigma$ grows with the feedback, except for the strongest feedback. The increase of $\sigma$ with the feedback is consistent with previous observations by Hong and Shore~\cite{hong_ol_2005} that found an increase of the relative mean dropout amplitude, $(\left<P_{max}\right>-\left<P_{min}\right>)/\left<P_{max}\right>$, with the feedback strength (here $\left<P_{max}\right>$ and $\left<P_{min}\right>$ are the mean levels between which the intensity falls during the dropouts). In Fig.~\ref{fig:3}(b) we note that when the current is normalized to the threshold of the laser with feedback, $\mathrm{I_{th}}$, the shape of the curve $\sigma$ vs. $\mathrm{I}/\mathrm{I_{th}}$ is very similar for all the feedback strengths, except for the highest one. This also occurs for the plot of $S$ and $K$  vs. $\mathrm{I}/\mathrm{I_{th}}$ (not shown).

Complementing the analysis of the intensity distribution, we next analyze the ordinal probabilities computed from the inter-dropout intervals (IDIs). As shown in Fig.~\ref{fig:1}(b), the threshold -1.5$\sigma$ allows to detect, in each time-series, a large number of dropouts. A clear boundary can be observed between the noisy and LFF regimes (that is fully consistent with the boundary detected with the analysis of the intensity distribution); however, the number of  dropouts varies smoothly during the LFF -- CC transition and in this figure the two regimes can not be quantitatively distinguished.

\begin{figure}
\centering
\includegraphics[width=\columnwidth]{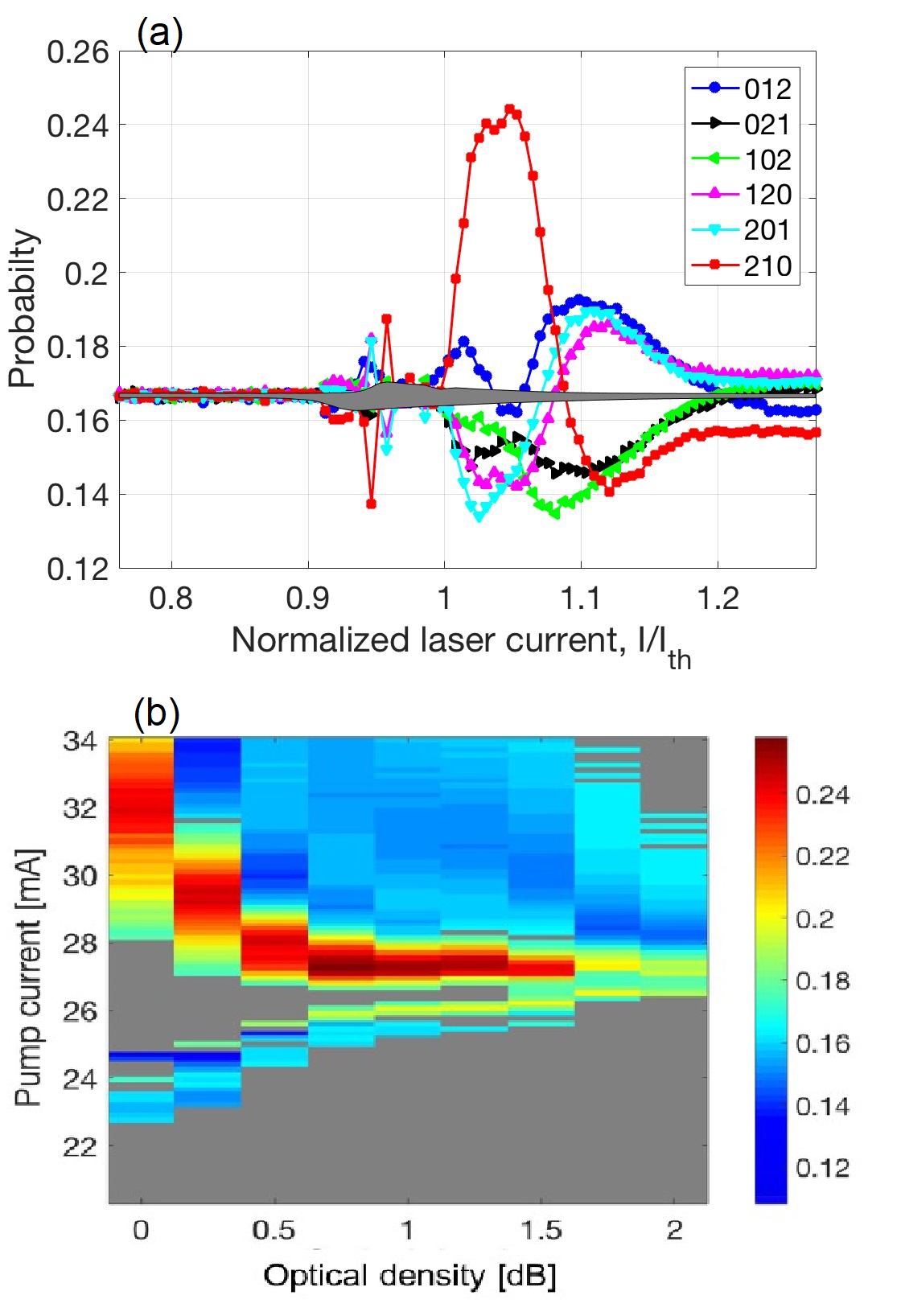}
\label{fig:4}
\caption{Ordinal analysis of experimental sequences of inter-dropout-intervals. (a) Probabilities of the six $D = 3$ ordinal patterns vs. the normalized pump current, $\mathrm{I}/\mathrm{I_{th}}$ for a constant feedback strength (NDF OD=0.4). The gray region indicates the interval of probability values that are consistent with the uniform distribution~\cite{gray_region}. (b) Probability of pattern 210 vs. the laser current and feedback strength. In the red/blue regions $P(210)$ is significantly higher/lower than expected if the patterns are equally probable; in the gray region $P(210)$ is within the interval of probability values that are consistent $P(210)=1/6$.%
}
\end{figure}

Figure~\ref{fig:4}(a) displays the six ordinal probabilities when the laser current is varied while the feedback strength is kept constant. The gray region represents probability values that are consistent with equally probable ordinal patterns~\cite{gray_region}. The variation of the six probabilities is consistent with previous observations~\cite{aragoneses_2014,quintero_sci_rep_2016}. The onset of the LFF regime is identified as the lowest pump current at which at least one probability is outside the gray region. This captures the emergence of temporal correlations associated to the first regime transition: from noise to LFFs. The onset of the CC regime is identified as the pump current value where the probability, $P$, of pattern 210 (i.e., three consecutive intervals being increasingly shorter) is maximum, because above this value, the gradual decrease of $P(210)$ uncovers the gradual disappearance of temporal correlations. 

Figure~\ref{fig:4}(b) displays $P(210)$ in color code  vs. the laser current and feedback strength. We chose this pattern because it is, in general, the one that differs the most from the 1/6 value expected for equally probable patterns \cite{sorrentino_2015}. 

In Fig.~\ref{fig:4}(b) we note that the red area where $P(210)$ is maximum (and thus indicates the LFF--CC boundary) moves, with increasing feedback, to higher values of the laser current. We also note that the two criteria employed to identify the laser current at which the LFF--CC transition occurs --the maximum of the intensity standard deviation, $\sigma$, and the maximum of probability of pattern 210, $P(210)$-- give consistent results, regardless of the feedback strength. 

In Fig.~\ref{fig:4}(b) the gray areas represent parameter regions where $P(210)$ is consistent with 1/6~\cite{gray_region}.  Comparing Figs.~\ref{fig:1}(a) and ~\ref{fig:4}(b) we note that the noise region in Fig.~\ref{fig:1}(a) is a gray region in Fig.~\ref{fig:4}(b), as expected for  uncorrelated fluctuations. In Fig.~\ref{fig:4}(b) we also note two gray regions, one that corresponds in Fig.~\ref{fig:1}(a) to LFFs, and the other, which corresponds to coherence collapse. In the gray LFF region, an inspection of the intensity time series reveals that the dropouts are regular, with well defined periodicity, and in this region all ordinal probabilities are $\sim 1/6$ [see Fig.~\ref{fig:4}(a)]. In the gray CC region in the top-right corner of Fig.~\ref{fig:4}(b) (for high current and high optical density of the neutral density filter, i.e., weak feedback strength) the intensity dynamics resembles noisy fluctuations and the ordinal probabilities are close to (or within the interval of values consistent with) 1/6. In this region $P(210)$ does not identify clear signatures of determinism in the sequence of inter-dropout intervals.

\section{\label{sec:numerical} Model and numerical results}

To demonstrate the robustness of the experimental observations we performed simulations of the Lang and Kobayashi (LK) rate equations \cite{lk_model} for the slowly varying complex electric field $E$ and the carrier density $N$. The model equations are: 
\begin{eqnarray} 
 \frac{dE}{dt} &=& k(1+\alpha)(G-1)E+\eta E(t-\tau)e^{-i\omega_{0}\tau} \nonumber \\
 &&+\sqrt{2\beta_{sp}}\;\xi\\ \frac{dN}{dt} &=& \frac{1}{\tau_{N}}\left(\mu-N-G \left| E \right| ^{2}\right)
 \end{eqnarray}
where $k$ is the cavity losses, $\alpha$ is the linewidth enhancement factor, $\tau_N$ is the carrier lifetime, $G = N/(1+\varepsilon \left| E \right|^{2})$ is the optical gain (with $\varepsilon$ a saturation coefficient). $\tau$ is the feedback delay time, $\tau=2L/c$, with $L$ being the length of the external cavity and $c$ the speed of light. $\omega_0$ is the solitary laser frequency, $\omega_0 \tau$ is the feedback phase, $\beta_{sp}$ is the noise strength, representing spontaneous emission, and $\xi$ is a Gaussian distribution with zero mean and unit variance. 

$\mu$ is the pump current parameter, which is proportional to the experimental control parameter, the laser current $I$. At the threshold of the solitary laser, $\mu$ is equal to the normalized current, $\mathrm{I}/\mathrm{I_{th,sol}}$, both being equal to~1~\cite{giacomelli}. $\eta$ is the feedback coefficient and $\eta^2$ is the feedback strength, which is inversely proportional to the optical density (OD) of neutral density filter (NDF) used in the experiments. 

The model equations were simulated with parameters as in \cite{quintero_sci_rep_2016}, where a qualitative good agreement model--experiments was reported: $\alpha=4.0$, $k=300$ ns$^{-1}$, $\tau_N =1$ ns,  $\varepsilon=0.01$, $\beta_{sp} = 10^{-4}$ ns$^{-1}$ and $\omega_0 \tau$ is a random value in (0,$2\pi$). To fit the experimental situation, $\tau=5$ ns.

For these parameters the LFFs are a transient dynamics with a duration that increases with the pump current and decreases with the feedback strength~\cite{torcini2006low,zamora2010transient}. 
In order to simulate the finite bandwidth of the experimental detection system, a simple filter is applied (a moving average over a time-window of $\Delta t=5$ ns). The influence of $\Delta t$ is discussed below. We use a simple moving window because this is sufficient to obtain a reasonable, but only qualitative, agreement with the experimental observations. As the LK model is well established, we speculate that a more advanced filtering method (that more precisely mimics the high-pass filtering performed by the detection system) will result in an improved quantitative agreement. In order to generate a sufficiently large number of dropouts, for each set of parameters (pump current, feedback strength) 20 trajectories of 100 $\mu$s each were generated from random initial conditions. 

\begin{figure}
\centering
  \includegraphics[width=.475\textwidth]{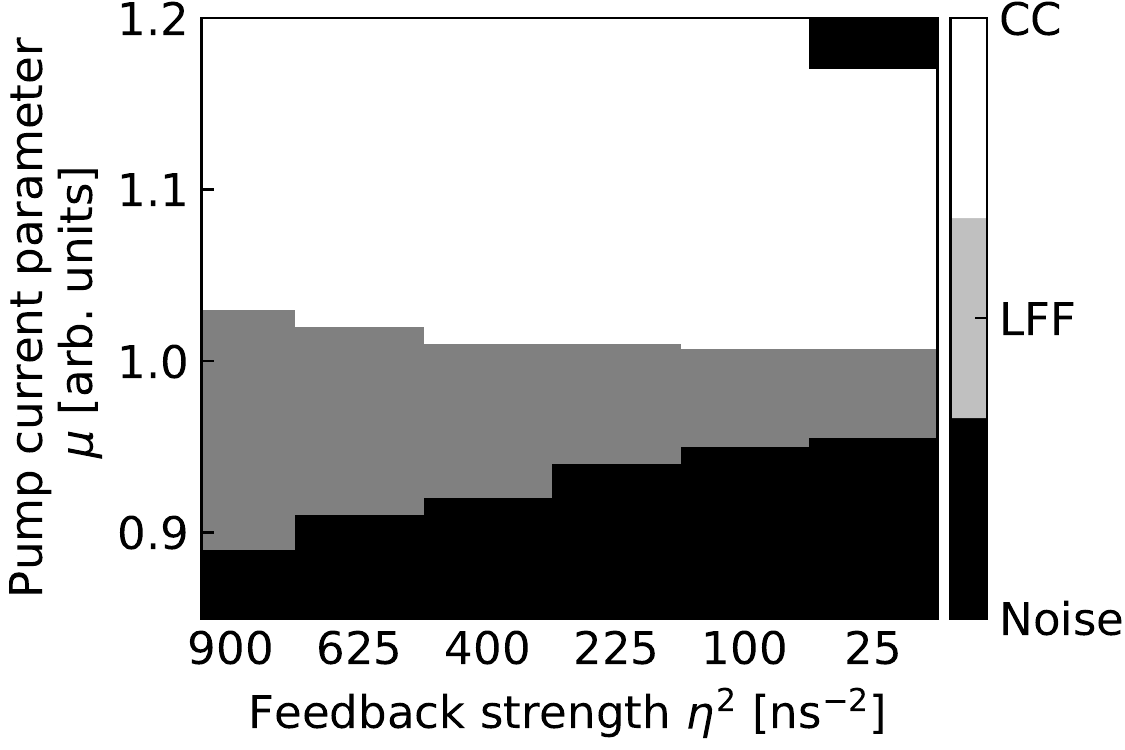}
\label{fig:mapa_numerico}
\caption{Numerical map that is obtained by using the same criteria as in Fig.~\ref{fig:1}(b) to classify the simulated intensity time series as either noise, LFF and CC. }  
\end{figure}

The regions of noise--LFF--CC dynamics are identified by using the same criteria as in the experimental data. Figure~\ref{fig:mapa_numerico} displays the obtained map, and we can note a good qualitative agreement with the experimental map displayed in Fig.~\ref{fig:1}(a).

\begin{figure}
\centering
\includegraphics[width=\columnwidth]{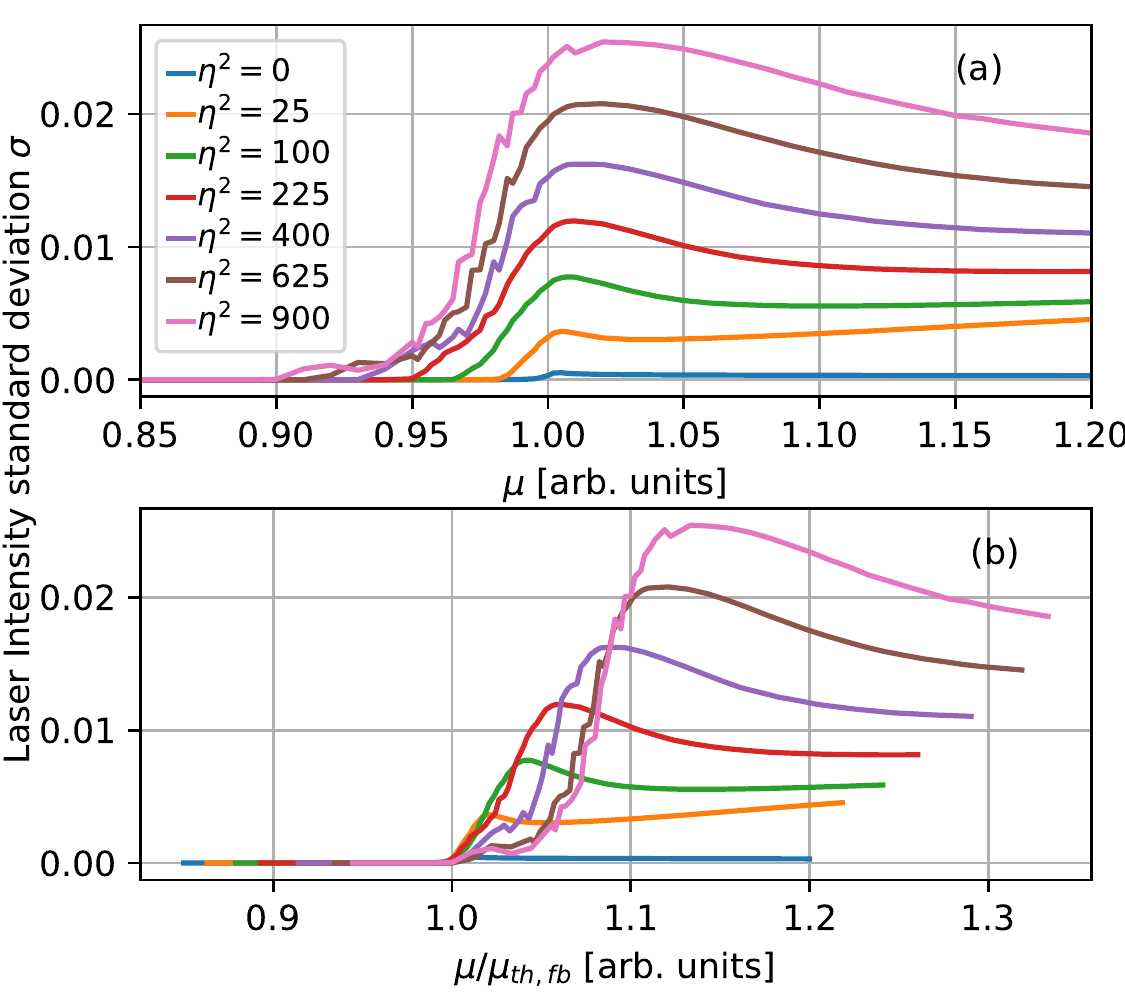}
\label{fig:sigma_numerico}
\caption{(a) Standard deviation of the simulated intensity time series vs. the laser current parameter, $\mu$, for various feedback strengths. In panel (b) $\mu$ is normalized to the threshold of the laser with feedback, $\mu_{th,fb}$. }
\end{figure}

\begin{figure}
\centering
\includegraphics[width=\columnwidth]{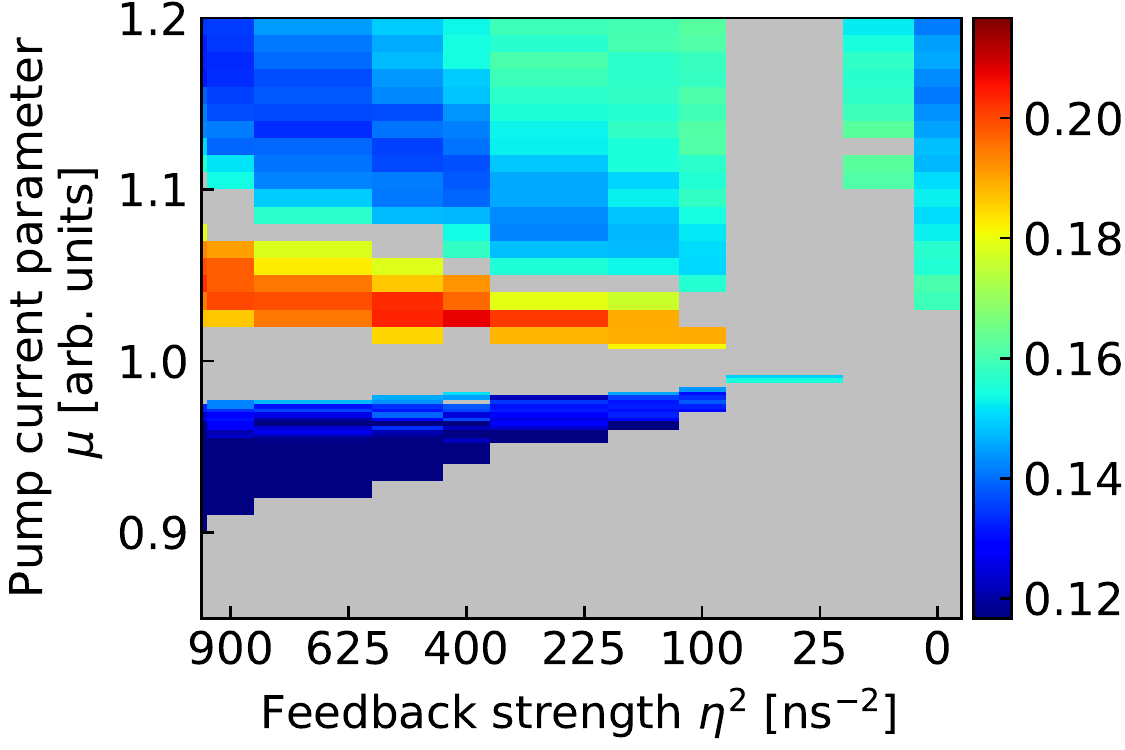}
\label{fig:p210num}
\caption{Probability of the ordinal pattern 210, computed from the simulated intensity time series. We note a reasonably good agreement with the experimental map, Fig.~\ref{fig:4}(b).}
\end{figure}

Figure~\ref{fig:sigma_numerico} displays the standard deviation of the simulated intensity time series, $\sigma$, vs. the pump current parameter, $\mu$, for different feedback strengths. We note that the shape of the curve $\sigma$ vs. $\mu$ is in good qualitative agreement with the one observed experimentally in Fig.~\ref{fig:3}. Figure~\ref{fig:p210num} displays $P(210)$ as a function of $\mu$ and $\eta^2$ and here again we see a reasonably good agreement with the experimental map, Fig.~\ref{fig:4}(b). 

\begin{figure}
\centering
  \includegraphics[width=0.475\textwidth]{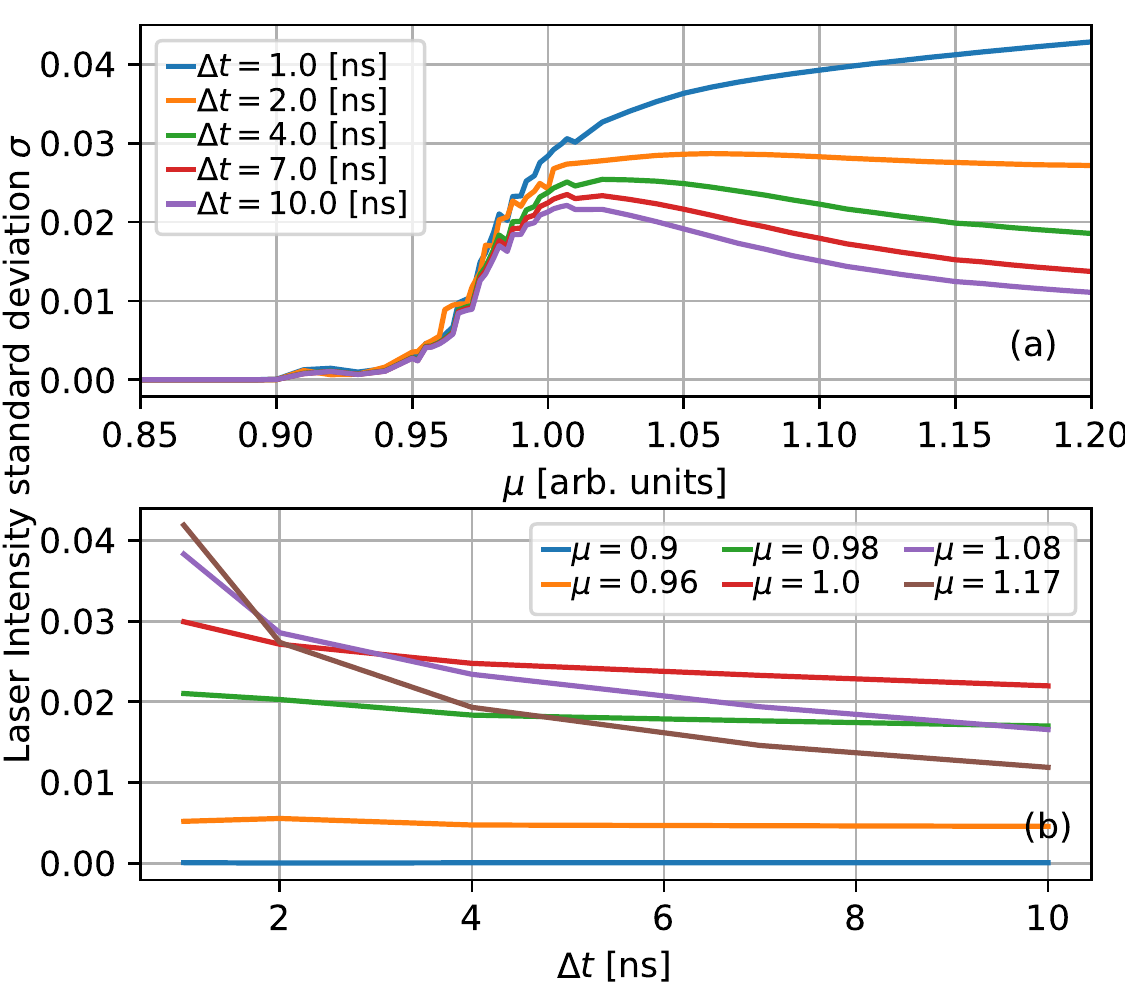}
  \label{fig:role_resolution_num}
  \caption{(a) Standard deviation of the simulated intensity time series as a function of the laser current parameter, $\mu$, for various values of the length of the moving window, $\Delta t$, that mimics the experimental detection bandwidth. (b) Standard deviation of the simulated intensity time series vs. $\Delta t$ for various values of $\mu$. In both panels $\eta=30$~ns$^{-1}$.}
\end{figure}

We conclude the analysis by discussing the influence of the experimental sampling rate, which, as explained before, was mimicked by averaging the simulated intensity time series over a moving window of size $\Delta t$. Figure~\ref{fig:role_resolution_num}(a) displays $\sigma$ vs. $\mu$ for various values of $\Delta t$. We note that for low $\mu$ (in the noise or in the LFF regimes) $\sigma$ remains rather unaffected by $\Delta t$, while for high $\mu$ (in the CC regime) $\sigma$ decreases with $\Delta t$. These two qualitatively different behaviors are shown in detail in panel (b). 

\begin{figure}
\centering
  \includegraphics[width=.55\textwidth]{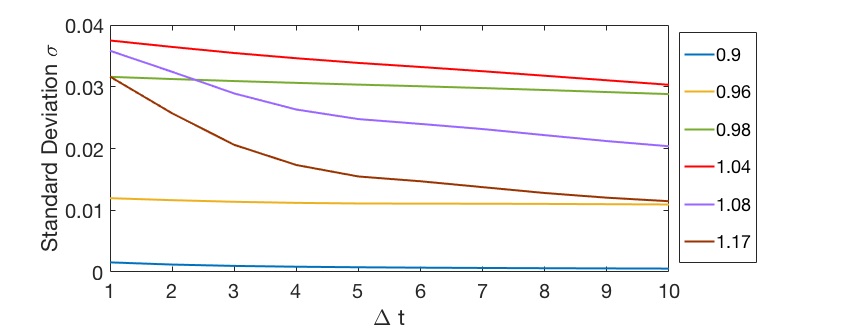}
  \caption{Experimental standard deviation of the intensity time series vs. $\Delta t$ for various values of the laser current, normalized to the solitary laser threshold. The optical density of the NDF is 0.6 [dB].}
  \label{fig:exp_Dt}
\end{figure}

This observation can be used to determine, from a single simulated intensity time series, if parameters are in the CC regime or not, without the need of comparing the value of $\sigma$ of the time series with the value of $\sigma$ of two ``neighboring'' time series. By plotting the value of $\sigma$ vs. $\Delta t$ we can determine if parameters are in the CC regime ($\sigma$ decreases with $\Delta t$) or not ($\sigma$ remains nearly constant when $\Delta t$ is varied). Figure~\ref{fig:exp_Dt} displays the result of the analysis of the experimental intensity time series, when the standard deviation is computed after averaging the intensity data values in a moving window of length $\Delta t$. A clear qualitative agreement is observed with the simulations presented in Fig.~\ref{fig:role_resolution_num}(b). The order of magnitude difference between numerical simulations and experiments are due to the simple moving averaging technique applied to the numerical data in order to mimic the experiments. We might obtain a better qualitative agreement by using a more precise high-pass filter detection system.

\section{\label{sec:con}Conclusions and discussion}

We have analyzed different dynamical regimes in the output of a semiconductor laser with optical feedback, as a function of the laser current and the feedback strength. We have shown that the shape of the probability distribution of intensity values (characterized by the standard deviation, the skewness and the kurtosis) complemented by ordinal analysis of the inter-dropout time intervals, allow to quantitatively distinguish among noisy intensity fluctuations, low-frequency fluctuations and coherence collapse. In particular we found that the two criteria employed to identify the laser current at which the LFF--CC transition occurs (the maximum of the intensity standard deviation, and the maximum of probability of pattern 210) give consistent results, for all the range of feedback strengths studied (2.6\% to 18.15\% threshold reduction). Simulations of the well-known time-delayed Lang-Kobayashi model were found to be in good qualitative agreement with the observations.

It is interesting to compare our results with the results of Toomey and Kane~\cite{deb_oe_2014}, that applied ordinal analysis to raw intensity time-series (without applying a threshold to detect the dropouts), and mapped the permutation entropy (that is Shannon's entropy computed from the ordinal probabilities) as a function of the laser current and the feedback strength. Low entropy regions were interpreted as due to the emergence of well-defined periodicity at the external cavity frequency, while high entropy regions were associated to either noise or coherence collapse [Fig. 2(b) of Ref.~\cite{deb_oe_2014}]. Our results are consistent with those of Toomey and Kane~\cite{deb_oe_2014} and also allow to quantitatively distinguish between the different regimes. 

As future work it would be interesting to apply these analysis tools to the dynamics induced by filtered optical feedback in a high power multi-mode laser. By mapping experimentally the dynamics as a function of the feedback strength and pump current, Baladi et al.~\cite{min_2016} recently found two distinct types of LFF regimes, distinguished by either intensity dropouts or jump-ups. It will also be interesting to analyze the so-called short cavity regime (where the feedback delay time is shorter than the solitary relaxation oscillation period of the laser), and to compare the maps obtained with the diagnostic tools used here, with the maps presented by Karsaklian Dal Bosco et al.~\cite{uchida_2017}, where LFF and CC regimes were distinguished by considerations based on the amplitude of the peaks in the RF spectra. 

\section*{Acknowledgement}
This work was supported in part by Spanish MINECO/FEDER (FIS2015-66503-C3-2-P) and ITN NETT (FP7 289146). C. M. also acknowledges partial support from ICREA ACADEMIA, Generalitat de Catalunya.


\begin{thebibliography}{99}

\bibitem{uchida_nat_phot_2008} A. Uchida, et al. 
Nat. Phot. \textbf{2}, 728–732 (2008).

\bibitem{donati_2012} S. Donati, 
Laser and Phot. Rev. \textbf{6}, 393–417 (2012).

\bibitem{uchida_chaos_2012} S. Sunada, T. Harayama, P. Davis, et al.,
Chaos \textbf{22}, 047513 (2012).

\bibitem{ingo_nat_comm_2013} D. Brunner, M. C. Soriano, C. R. Mirasso, and I. Fischer,  
Nat. Comm. \textbf{4}, 1364 (2013).

\bibitem{marc_nat_phot_2015} M. Sciamanna and K. A. Shore, 
Nat. Phot. \textbf{9}, 151 (2015).

\bibitem{rontani_sci_rep_2016} D. Rontani, D. Choi, C. Y. Chang, et al.,
Sci. Rep. 6, 35206 (2016).

\bibitem{masoller_chaos_1997} C. Masoller, 
Chaos 7, 455-462 (1997).

\bibitem{parlitz_pre_1998} V. Ahlers, U. Parlitz and W. Lauterborn, 
Phys. Rev. E \textbf{58}, 7208-7213 (1998).

\bibitem{vicente_jqe_2005} R. Vicente, J. Dauden, P. Colet and R. Toral, 
IEEE J. Quantum Electron. \textbf{41}, 541–548 (2005).

\bibitem{yanchuk_prl_2014} S. Yanchuk and G. Giacomelli, Phys. Rev. Lett. \textbf{112}, 174103 (2014).

\bibitem{tiana_pra_2010} J. Tiana-Alsina, M. C. Torrent, O. A. Rosso, C. Masoller and J. Garcia-Ojalvo, 
Phys. Rev. A \textbf{82}, 013819 (2010).

\bibitem{soriano_jqe_2011} M. C. Soriano, L. Zunino, O. A. Rosso et al.,
IEEE J. Quantum Electron. \textbf{47}, 252-261 (2011). 

\bibitem{zunino_jqe_2011} L. Zunino, O. A. Rosso and M. C. Soriano,
IEEE J. Sel. Top. Quantum Electron. \textbf{17}, 1250 (2011). 

\bibitem{nico_pre_2011} N. Rubido, J. Tiana-Alsina, M. C. Torrent, et al.,
Phys. Rev. E \textbf{84}, 026202 (2011).

\bibitem{aragoneses_2014} A. Aragoneses, S. Perrone, T. Sorrentino, et al.,
Sci. Rep. \textbf{4}, 4696 (2014). 


\bibitem{deb_oe_2014} J. P. Toomey and D. M. Kane, 
Opt. Express \textbf{22}, 1713 (2014).

\bibitem{porte_autocorrel_pre_2014} X. Porte, O. D'Huys, T. Jungling, et al.,
Phys. Rev. E  90, 052911 (2014).

\bibitem{porte_similarity_pra_2014} X. Porte, M. C. Soriano and I. Fischer, 
Phys. Rev. A 89, 023822 (2014).

\bibitem{masoller_njp_2015} C. Masoller, Y. Hong, S. Ayad, et al.,
New J. of Phys. \textbf{17}, 023068 (2015). 

\bibitem{masoller_plos_2016} B. L. Lan and C. Masoller, 
PLoS ONE 11, e0150027 (2016).

\bibitem{quintero_sci_rep_2016} C. Quintero-Quiroz, J. Tiana-Alsina, J. Roma, M. C. Torrent, and C. Masoller,
Sci. Rep. \textbf{6} 37510 (2016). 

\bibitem{lenstra_1985} D. Lenstra, B. Verbeek and A. Den Boef, 
IEEE J. Quantum Electron. \textbf{21}, 674 (1985).

\bibitem{feedback_regimes_1986} R. Tkach and A. Chraplyvy, J. Lightwave Technol. \textbf{4}, 1656 (1986).

\bibitem{shore_ol_1994}  L. N. Langley, K. A. Shore and J. Mork,
Opt. Lett. 19, 2137 (1994).

\bibitem{ingo_coex_1998} T. Heil, I. Fischer and W. Elssaber, 
Phys. Rev. A \textbf{58}, R2672 (1998).

\bibitem{video} An experimental video of the transition is available at \url{https://www.youtube.com/watch?v=nltBQG_IIWQ&feature=youtu.be}

\bibitem{bandt2002permutation} C. Bandt and B. Pompe, 
Phys. Rev. Lett. \textbf{88}, 174102 (2002).

\bibitem{uchida_2017} A. Karsaklian Dal Bosco, S. Ohara, N. Sato, et al.,
IEEE Phot. Journal \textbf{9}, 6600512 (2017).

\bibitem{ingo_2001} T. Heil, I. Fischer, W. Elsasser, and A. Gavrielides, 
Phys. Rev. Lett. \textbf{87},  243901 (2001).

\bibitem{marc_2004} A. Tabaka, K. Panajotov, I. Veretennicoff, and M. Sciamanna,
Phys. Rev E \textbf{70}, 036211 (2004).

\bibitem{pepe_2013} J. A. Reinoso, J. Zamora-Munt, and C. Masoller, 
Phys. Rev. E \textbf{87}, 062913 (2013).

\bibitem{lk_model} R. Lang and K. Kobayashi, 
IEEE J. Quantum Electron. \textbf{16}, 347 (1980).

\bibitem{ingo_prl_1996} I. Fischer, G. H. M. van Tartwijk, A. M. Levine, et al., Phys. Rev. Lett. \textbf{76}, 220 (1996).
\bibitem{gray_region} The gray region represents the interval of probability values which are consistent with the uniform distribution: $p \pm 3 \sigma_p$ (confidence level of 99.5\%), where $p = 1/6$ and $\sigma_p = \sqrt{p(1 − p)/N}$, with $N$ being the number of inter-dropout intervals in the time-series.
\bibitem{hong_ol_2005} Y. Hong and K. A. Shore Opt. Lett. \textbf{30}, 3332 (2005). 
\bibitem{sorrentino_2015} T. Sorrentino, C. Quintero-Quiroz, M. C. Torrent, and C. Masoller IEEE J. Sel. Top. Quantum Electron. \textbf{21}, 1801107 (2015).
\bibitem{giacomelli} S. Barland, P. Spinicelli, G. Giacomelli, and F. Marin, IEEE J. Quantum Electron. \textbf{41}, 1235 (2005).
\bibitem{torcini2006low} A. Torcini, S. Barland, G. Giacomelli,  and F. Marin, Phys. Rev. A \textbf{74}, 063801 (2006).
\bibitem{zamora2010transient} J. Zamora-Munt, C. Masoller,  and J. Garc\'ia-Ojalvo, Phys. Rev. A \textbf{81}, 033820 (2010).
\bibitem{min_2016} F. Baladi, M. W. Lee, J-R Burie et al.,
Opt. Lett. \textbf{41}, 2950 (2016). 
\end{thebibliography}
\end{document}